\documentclass[showpacs,twocolumn,prb]{revtex4}
\usepackage{graphicx}
\newcommand{\be}{\begin{equation}}
\newcommand{\ee}{\end{equation}}
\newcommand{\bea}{\begin{eqnarray}}
\newcommand{\eea}{\end{eqnarray}}
\newcommand{\bwt}{\begin{widetext}}
\newcommand{\ewt}{\end{widetext}}

\begin{document}
\title{Temperature Dependence of the Conductivity Sum Rule in the
Normal State due to Inelastic Scattering}
\author{L. Benfatto$^1$, J.P.Carbotte$^2$, and F. Marsiglio$^3$}
\affiliation{$^1$SMC-INFM-CNR and Department of Physics, University
of Rome ``La Sapienza'', Piazzale Aldo Moro 5, 00185, Rome, Italy\\
$^2$Department of Physics and Astronomy, McMaster University,
        Hamilton, Ontario, Canada, L8S 4M1 \\
$^3$Department of Physics, University of Alberta,
Edmonton, Alberta, Canada, T6G~2J1 \\
$^3$DPMC, Universit\'e de Gen\`{e}ve, 24 Quai Ernest-Ansermet,
CH-1211 Gen\`{e}ve 4, Switzerland, \\
$^3$National Institute for Nanotechnology, National Research Council
of Canada, Edmonton, Alberta, Canada, T6G~2V4}

\begin{abstract}
We examine the temperature dependence of the optical sum rule in the
normal state due to interactions. To be concrete we adopt a weak
coupling approach which uses an electron-boson exchange model to
describe inelastic scattering of the electrons with a boson, in the
Migdal approximation. While a number of recent works attribute the
temperature dependence in the normal state to that which arises in a
Sommerfeld expansion, we show that in a wide parameter regime this
contribution can be quite small. Instead, most of the temperature
dependence arises from the zeroth order term in the `expansion',
through the temperature dependence of the spectral function, and the
interaction parameters contained therein. For low boson frequencies
this circumstance causes a linear $T$-dependence in the sum rule. We
develop some analytical expressions and understanding of the
temperature dependence.

\end{abstract}

\date{\today}
\maketitle

\section{introduction}

In the last few years much attention has been focussed on the anomalous
temperature dependence of the optical sum rule
\cite{basov99,molegraaf02,santander-syro03,homes,boris,ortolani,hwang,bontemps06}.
In particular, in a number of superconductors in the underdoped and
optimally doped regime the sum rule exhibits a temperature dependence which
is anomalous with respect to that expected from a conventional BCS
condensation. The origin of this behaviour has been the subject of a number
of theoretical explanations
\cite{hirsch92,hirsch00,norman02,karakozov02,schachinger05,marsiglio06}.
It is probably fair to say that there is no consensus concerning the origin
of the observed anomaly below the superconducting critical temperature,
$T_c$.

Perhaps equally intriguing is the normal state temperature
dependence.\cite{knigavko04,benfatto05,knigavko05,toschi05,benfatto06} Most
often the value of the sum rule is plotted versus $T^2$, where $T$ is the
absolute temperature, and a linear relationship is observed.
\cite{molegraaf02,santander-syro03,ortolani} This appears natural as a
straightforward Sommerfeld expansion produces a finite temperature
correction which is quadratic in temperature.  However, several researchers
\cite{ortolani,benfatto05,toschi05} have noticed that the observed
temperature dependence is too strong compared with what one would expect
with realistic bandwidths in the cuprate materials (for a quantitative
analysis of experimental data in cuprates see
Ref. \onlinecite{benfatto06}).  Moreover, there are clear deviations from
$T^2$ dependence, \cite{santander-syro03} along with some indications of a
linear temperature variation of the spectral weight \cite{hwang,bontemps06}
so that one is left with the question: within a Fermi liquid approach, what
governs the temperature dependence of the optical sum rule in the normal
state?

This is the question we address in this paper. To be concrete we use
an exchange boson model (Migdal approximation for phonons), which is
a weak-coupling approach. This approach is believed to be applicable
to phonons, whose possible relevance to the cuprates has been
questioned in the last few years \cite{lanzara01}. It is more
suspect from a theoretical point of view for magnon exchange;
nonetheless we simply adopt the Migdal approximation, with
justification provided by some researchers \cite{chubukov05}. This
type of calculation has also recently been carried out by Karakozov
and Maksimov \cite{karakozov05}. We reinforce their (and our) main
message: a strong temperature dependence of the optical sum rule can
arise due to interactions. We examine in further detail the
dependence on the various parameters of the model. In particular, we
show that the finite width of the band, which is crucial for
discussing the sum-rule temperature dependence, introduces new
features in the electron self-energy, which are usually overlooked
in the infinite-bandwidth approximation. Moreover, we show that the
sum-rule temperature dependence cannot be attributed merely to the
temperature dependence of the imaginary part of the self energy.
Indeed, the real part enters in a critical way.

The outline of the paper is as follows. In the next section we
outline the formalism, both on the real and imaginary axes. The
former is more appropriate for analytical work, whereas the latter
is ideal for numerical work. We show results for a wide variety of
parameters, which clearly shows how the temperature dependence can
go from quadratic to linear in temperature, depending on the boson
frequency scale. We also outline how the Sommerfeld contribution can
be determined (even in the presence of interactions). It is small
always since we use the Fermi energy as a dominant energy scale. We
also show how a large but non-infinite energy scale actually
qualitatively changes `low energy' results. In section III we adopt
a sequence of approximations to illustrate precisely what is
required to achieve qualitative and quantitative agreement with the
numerical results. We find that the real part of the single particle
self energy is required for a clear understanding of the temperature
dependence of the sum rule. Finally, we close with a summary.

\section{formalism}

The "single band" sum rule \cite{maldague77,hirsch00,norman02,benfatto05} can
be written
\begin{equation}
\int_{0}^{+\infty }d\nu \mathop{\rm Re} \left[ \sigma_{\rm xx} (\nu
)\right] = {\pi e^{2} \over 4\hbar^2} \biggl\{ \frac{4}{N}\sum_{k}
{\partial^2\epsilon_k \over \partial k_x^2} n_{k}. \biggr\}%
\label{sumrule}
\end{equation}
where $\epsilon_k$ is the tight-binding dispersion and $n_k$ is the
single spin momentum distribution function. In the case of two
dimensions with nearest neighbour hopping only, the quantity in
braces is given by the {\em negative} of the expectation value of
the kinetic energy
\begin{equation}%
K(T) = \frac{2}{N}\sum_{k} \epsilon_k n_{k}.%
\label{kinetic_energy}
\end{equation}%
We have explored elsewhere that the kinetic energy is a good monitor
of the sum rule, even in cases beyond nearest neighbour
hopping.\cite{benfatto05,vandermarel03}
Therefore in what follows we will use Eq. (\ref{kinetic_energy}), as
it is much simpler to calculate for a given model than the
left-hand-side of Eq. (\ref{sumrule}).

In general, the momentum distribution function is given by
\begin{eqnarray}
n_k & = & {1 \over \beta} \sum_{m = -\infty}^{+\infty}
G(k,i\omega_m) e^{i\omega_m0^+} = \nonumber\\
&=&{1 \over 2} + {1 \over \beta}
\sum_{m = -\infty}^{+\infty} {\rm Re} \ G(k,i\omega_m),%
\nonumber \\
& = & \int_{-\infty}^{+\infty}d\omega \ f(\omega) A(k,\omega)
\label{momentum_distribution} %
\end{eqnarray}
where $G(k,i\omega_m)$ is the single electron Matsubara Green
function with momentum $k$ and frequency $i\omega_m$. The
$i\omega_m$ are the Fermionic Matsubara frequencies, $i\omega_m =
\pi T (2m-1)$, where $m$ is an integer, $T$ is the temperature ($k_B
\equiv 1$), and $\beta \equiv 1/T$ is the inverse temperature. The
spectral function, $A(k,\omega) \equiv -{1 \over \pi} {\rm
Im}G(k,i\omega_m \rightarrow \omega + i\delta)$, is related to the
analytical continuation of the Matsubara Green function. Finally,
$f(\omega) \equiv 1/(e^{\beta \omega} + 1)$ is the Fermi-Dirac
distribution function. In the absence of interactions the spectral
function is a temperature-independent Dirac delta function, so that
$n_k\equiv f(\epsilon_k-\mu)$ and Eq.\ (\ref{kinetic_energy})
reduces to:
\begin{equation}
K_{\rm free}(T)=\frac{2}{N}\sum_k\epsilon_k f(\epsilon_k-\mu)
\label{non-int}
\end{equation}
It is then clear that the only temperature dependence of the sum
rule (\ref{non-int}) can be due to the "thermal smearing", i.e. the
effect contained within the Fermi-Dirac distribution function. A
Sommerfeld expansion then reveals this temperature dependence to be
quadratic in $T/W$, \cite{vandermarel03,benfatto05} where $W$ is the
electronic bandwidth; this parameter is usually small at
temperatures of interest (up to $200 \sim 300$ K). However, in the
general case of an interacting system Eq.\ (\ref{kinetic_energy})
can be written, by means of Eq.\ (\ref{momentum_distribution}), as
\begin{equation}
K(T)=2\int d\epsilon \ d\omega \ N(\epsilon)\epsilon
A(\epsilon,\omega) f(\omega) \label{int}
\end{equation}
where we converted the momentum sum to an energy integration in Eq.\
(\ref{kinetic_energy}).  In the presence of interactions, the Green
function $G$ is given by the Dyson equation,
\begin{equation}
G(k,z) = 1/\bigl[z - \epsilon_k - \Sigma(k,z)\bigr], %
\label{dyson}%
\end{equation}
where $z$ is anywhere in the complex plane, and $\Sigma(k,z)$ is the
electron self energy. This quantity, and consequently the spectral
function $A(\epsilon,\omega)$, will be in general temperature
dependent, and it is this temperature dependence that will manifest
itself in the total kinetic energy (\ref{int})(and therefore the
conductivity sum) that we wish to understand.\cite{remark1} Indeed,
using for simplicity a constant density of states in Eq.\
(\ref{int}), an application of the Sommerfeld expansion produces
\cite{ashcroft76}
\begin{eqnarray}
K(T)&=& K_0(T)+K_{somm}(T)=\nonumber\\
&=&K_0(T) + {\pi^2 \over 6} (k_BT)^2 {dH(\omega) \over
d\omega}|_{\omega=0}%
\label{koft}
\end{eqnarray}%
where the second term is what we refer to in what follows as the
Sommerfeld term (with chemical potential assumed to be zero, in the
middle of the band) and
\begin{equation}%
H(\omega) = {2 \over W} \int_{-W/2}^{+W/2} d\epsilon \ \epsilon
A(\epsilon,\omega).%
\label{ksomm}%
\end{equation}%
In the non-interacting case $H(\omega)$ reduces to $2\omega/W$, so
that the $T^2/W$ dependence of the Sommerfeld term is recovered, as
stated above.  In the interacting case instead $dH(\omega)/d\omega$ is
affected by the quasiparticle renormalization \cite{toschi05} (see
Eq.\ (\ref{ksomm1}) below) and can also bring additional temperature
dependence to the $T^2$ factor of Eq.\ (\ref{koft}). The first term
$K_0(T)$ in Eq.\ (\ref{koft}) reduces to the constant value $K(T=0)$
in the absence on interactions, see Eq.\ (\ref{non-int}). However, in
the general case it is given by
\begin{equation}
K_0(T) = \int_{-\infty}^0 d\omega \ H(\omega) = \int_{-\infty}^0
d\omega \ {2 \over W} \int_{-W/2}^{+W/2} d\epsilon \ \epsilon
A(\epsilon,\omega).%
\label{k0}
\end{equation}%
The key observation, also made (independently) by Karakozov and
Maksimov \cite{karakozov05} is that Eq. (\ref{k0}) can have a far more
significant temperature dependence than the Sommerfeld
contribution. Moreover, as we shall investigate here, due to this
effect the temperature dependence of the kinetic energy can deviate
from the non-interacting case.

As mentioned in the introduction, this work will focus on
``conventional'' interactions, modelled after the electron-phonon
interaction, as described by an Einstein oscillator with frequency
$\omega_E$ and coupling strength $\lambda$.  In the
non-selfconsistent Migdal approximation, the self energy is given
by:
\begin{equation}
\Sigma(i\omega_m) = {W \over N\beta} \sum_{k^\prime, m^\prime}
\lambda(i\omega_{m^\prime} - i\omega_m)
G_0(k^\prime,i\omega_{m^\prime}),%
\label{migdal}%
\end{equation}%
where the non-selfconsistency is reflected in the fact that the
non-interacting electron Green function is used (hence the subscript
zero). This can be made self-consistent but the changes are minor
because of the large bandwidth we will always adopt (in other words
we are not pursuing the possibility of strong temperature dependence
arising from a low electronic energy scale). Here
\begin{equation}
\lambda(z) \equiv {\lambda \omega_E^2 \over \omega_E^2 - z^2},
\label{lambda}%
\end{equation}%
where $z$ is anywhere in the complex plane. The bandwidth $W$ is
included in Eq. (\ref{migdal}) since it is customary to have a
density of states factor in $\lambda$ (which, among other things,
makes it dimensionless), and so we use the `average' density of
states, $1/W$. One can readily evaluate Eq. (\ref{migdal}) for
various tight-binding models and use this in Eqs.
(\ref{dyson},\ref{momentum_distribution},\ref{kinetic_energy}).
However, we simplify further before presenting results and advise
the reader that such details make little difference in our conclusions.
To this end we assume a constant density of states for energies
$-W/2 < \epsilon < W/2$. Then the k-summation becomes a single
integral over energy, $\epsilon$. The result is
\begin{eqnarray}
\Sigma(i\omega_m) = {\lambda \omega_E \over 2} \int_{-W/2}^{+W/2}
d\epsilon \ \biggl({N(\omega_E) + f(\epsilon) \over i\omega_m +
\omega_E - \epsilon}  \nonumber\\
+{1 + N(\omega_E) - f(\epsilon) \over
i\omega_m -
\omega_E - \epsilon} \biggr)%
\label{migdala}%
\end{eqnarray}%
where $N$ and $f$ are the Bose and Fermi functions, respectively.
Eq. (\ref{migdala}) can be evaluated either at the Matsubara
frequencies $i\omega_m \equiv i\pi T(2m-1)$, with $m$ an integer, or
at the real frequencies attained through direct analytical
continuation, $i\omega_m \rightarrow \omega + i \delta$, with $i
\delta$ an infinitesimal imaginary part. To obtain reliable numerical
results the most straightforward approach is on the imaginary axis,
as we now describe.

\subsection{Imaginary axis formulation}
First, one evaluates the integrals in Eq. (\ref{migdala})
numerically. One can show that the result is pure imaginary (because
of the particle-hole symmetry), and so one can define a
``renormalization'' function through $\Sigma(i\omega_m) \equiv
i\omega_m (1 - Z(i\omega_m))$, where $Z(i\omega_m)$ is a real-valued
function. Next one returns to the kinetic energy; the momentum
dependence (i.e. the $\epsilon$ dependence) in Eq.
(\ref{kinetic_energy}) is simple enough (see the first line of Eq.
(\ref{momentum_distribution}) with $\epsilon$ dependence coming only
from the explicit dependence shown in Eq. (\ref{dyson})) to be done
analytically. The result is:
\begin{equation}%
K(T) = -{4 \over \beta}\sum_{m=1}^\infty \Biggl[1 - {\omega_m
Z(i\omega_m) \over W/2} {\rm tan}^{-1}\biggl({W/2 \over \omega_m
Z(i\omega_m)}\biggr) \Biggr].%
\label{kin1}%
\end{equation}%
This is easily summed provided asymptotics are used for the
contribution at high frequencies. Eq. (\ref{kin1}) with
$Z(i\omega_m)$ obtained from Eq. (\ref{migdala}) provides us with
our numerical results.
\begin{figure}[tp]
\begin{center}
\includegraphics[scale=0.4]{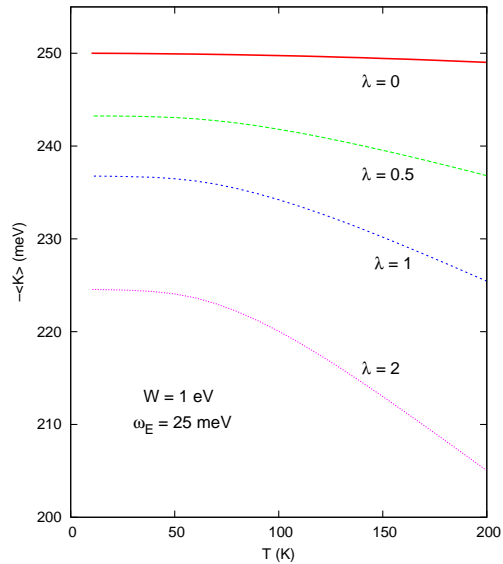}
\caption{ The negative of the kinetic energy vs. temperature for a
variety of coupling strengths. Note that the magnitude of the
temperature variation increases considerably with coupling strength,
and, as explained in the text, this {\em cannot} be attributed to
the impact of interactions on the Sommerfeld term.}
\end{center}
\end{figure}

It is difficult to develop an intuition for what temperature
dependence may emerge from various parameter choices. In particular
we have uncertainty about the energy scale of the boson. Before
developing some analytical approximations we therefore show some
results. Fig.~1 shows the (negative of the) kinetic energy vs.
temperature for several coupling strengths: $\lambda =0, 0.5, 1$ and
$2$ as labeled. We have used a consistent set of parameters for
these results: (i) a bandwidth $W=1$ eV, and (ii) a boson frequency
$\omega_E = 25$ meV. The first most noticeable trend is that the
absolute value of the kinetic energy decreases with increasing
coupling strength. This is expected, as coupling to a boson
'decoheres' the electrons. While the momentum distribution function
retains a discontinuity at the Fermi level (see, for example, Fig. 1
in Ref. \onlinecite{marsiglio06}), the overall distribution is
considerably more smeared than in the non-interacting case. Thus, in
the ground state, higher kinetic energy (i.e. lower absolute value)
states are occupied while low kinetic energy states remain
unoccupied. This sort of trend with coupling strength would be very
difficult to observe in the experiments, however, as there is no
simple way to compare two systems whose only difference is the
coupling strength. A more promising property to focus on is the
temperature dependence, which clearly varies as the coupling
strength increases. Before doing this we wish to make one important
remark concerning our parameter choices. Note that the bandwidth has
been chosen to be by far the largest energy scale in the problem.
This is generally the case in `conventional' metals. Interesting
effects can arise when the bandwidth is not so large compared to the
other energy scales (such as the boson frequency or the temperature
itself) \cite{dogan03,cappelluti03}. However, here we wish to
demonstrate that even in the `conventional' case a significant
temperature dependence of the kinetic energy appears due to
interactions.

\begin{figure}[tp]
\begin{center}
\includegraphics[scale=0.4]{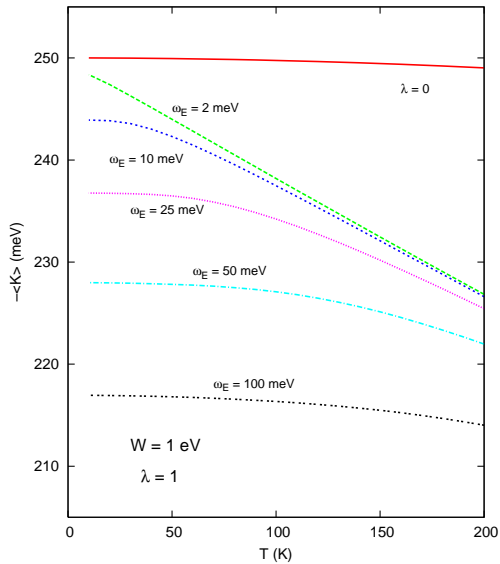}
\caption{ The negative of the kinetic energy vs. temperature for a
variety of boson frequencies. Note that with decreasing phonon
frequency not only is the magnitude of the temperature variation
increased, but it clearly crosses over from quadratic to linear in
temperature.}
\end{center}
\end{figure}

As we explained above, for the non-interacting case the deviation of
$K(T)$ from a constant is given by the $T^2/W$ Sommerfeld
contribution. \cite{benfatto05,benfatto06} The finite-temperature
corrections for the interacting cases shown in Fig.\ 1 are clearly not
simply quadratic, and furthermore the magnitude of the variation
increases substantially with coupling strength. Is this due to
alterations in the Sommerfeld contribution? The answer is most
definitively no. Indeed, even though the Sommerfeld correction (to be
further described below) changes by at most $65\%$ (at $200$ K) from
$\lambda = 0$ to $\lambda = 2$, this gives a correction of about $1.7$
meV (at $200$ K) for $\lambda=2$, whereas the deviation from the zero
temperature result is $\approx 20$ meV. Thus the bulk of the
temperature correction arises from other sources.

How does the energy scale of the Einstein frequency $\omega_E$
affect the temperature dependence? In Fig. 2 we show the (negative
of the) kinetic energy vs. temperature for several Einstein
frequencies with the same bandwidth as before, and with a moderate
coupling strength, $\lambda = 1$. It is clear that the energy scale
of the boson frequency has a profound effect both on the zero
temperature value and the temperature dependence itself. In
particular, for low boson frequencies the temperature dependence is
linear; in addition, as will be made clear below, the temperature
dependence comes almost entirely from $K_0(T)$, not the Sommerfeld
term. In contrast, for high boson frequency the temperature
dependence resembles that of the non-interacting case and a
significant fraction comes from the Sommerfeld term.

\subsection{Real axis formulation: Sommerfeld contribution}

\begin{figure}[htp]
\begin{center}
\includegraphics[scale=0.4]{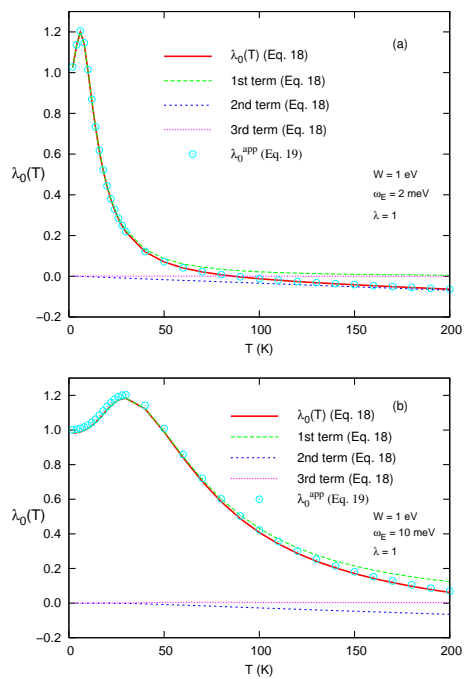}
\caption{ The mass enhancement parameter vs. temperature for (a)
$\omega_E = 2$ meV, and (b) $\omega_E = 10$ meV. Note the (very)
slight deterioration of the approximate result as the boson
frequency increases. Also note that for low boson frequency the mass
enhancement parameter becomes negative for high temperatures (but
`high' being defined relative to the boson, not the electronic
energy scale). Counterintuitively, the result for infinite bandwidth
(dashed green curve) becomes poorer as the boson energy scale {\em
decreases}.}
\end{center}
\end{figure}

To see how this significant temperature variation arises, let us go
back to the real-axis formulation (\ref{int})-(\ref{koft}) of the
kinetic energy and let us analyze the Sommerfeld contribution. To
obtain the self energy on the real axis a straightforward analytical
continuation of Eq. (\ref{migdala}) is required,
\begin{eqnarray}
\Sigma(\omega + i\delta) = {\lambda \omega_E \over 2}
\int_{-W/2}^{+W/2} d\epsilon \ \biggl({N(\omega_E) + f(\epsilon)
\over \omega + i\delta + \omega_E - \epsilon} + \nonumber\\
{1 + N(\omega_E) -
f(\epsilon) \over \omega + i\delta - \omega_E - \epsilon} \biggr).%
\label{migdalb}%
\end{eqnarray}%
Writing the real and imaginary parts as $\Sigma = \Sigma_1 +
i\Sigma_2$, it is easy to see that $\partial
\Sigma_2(\omega)/\partial \omega$ is zero at zero frequency.
Furthermore, $\Sigma_1(\omega=0) = 0$. Defining, as  usual,
$\partial \Sigma_1(\omega)/\partial \omega |_{\omega=0} \equiv
-\lambda_0(T)$, then we obtain,
\begin{equation}%
{\partial A(\epsilon,\omega) \over \partial \omega} |_{\omega=0} =
-{2 \over \pi} \epsilon (1 + \lambda_0(T)) {\Sigma_2(T,\omega=0)
\over [\epsilon^2 + \Sigma_2^2(T,\omega=0)]^2},%
\label{ksomm1}%
\end{equation}%
Substitution into the second term of Eq. (\ref{koft}) yields, for
the Sommerfeld contribution,
\begin{eqnarray}%
K_{\rm Somm} = {\pi \over 3} (1 + \lambda_0(T)) \biggl({T \over
W/2}\biggr)^2 {W \over 2}\times \nonumber\\
 \Biggl({\rm tan}^{-1}\biggl({W/2 \over
|\Sigma_2(T)|}\biggr)
 - {{|\Sigma_2(T)| \over W/2} \over 1 +
\bigl({|\Sigma_2(T)| \over W/2}\bigr)^2} \Biggr).
\label{ksomm2}%
\end{eqnarray}%
Here, the zero frequency imaginary part of the self energy (denoted simply
by $\Sigma_2(T)$) is required; for a constant density of states the (exact)
result is given by the result known as being a good approximation for the
infinite bandwidth case \cite{allen82},
\begin{equation}%
\Sigma_2(T) = -\pi \lambda \omega_E/{\rm sinh}(\beta \omega_E).%
\label{sig20}%
\end{equation}%
At low temperatures this is zero, while at high temperatures (with
respect to the boson frequency, $\omega_E$), this is linear in
temperature.

The real part of the self energy, required to determine
$\lambda_0(T)$, is more difficult. For $\omega_E < W/2$ (the
physically relevant case) we obtain
\begin{eqnarray}%
\lambda_0(T) = \lambda \Biggl( & & 4 \pi T \sum_{m=1}^\infty
{\omega_m \omega_E^2 \over [\omega_m^2 + \omega_E^2]^2} \Bigl[ 1 -
{2 \over \pi} {\rm tan}^{-1}\bigl({\omega_m \over W/2}\bigr) \Bigr]%
\nonumber \\
& & -{2 \over {\rm sinh}(\beta \omega_E)} {\omega_E \over W/2}{1
\over 1 - \bigl({\omega_E \over W/2}\bigr)^2}%
\nonumber \\
& & + {\omega_E \over 2T}f(\omega_E)[1 - f(\omega_E)] {\rm
ln}\Biggl|{1 + {\omega_E \over W/2} \over 1 - {\omega_E \over
W/2}}\Biggr|%
\Biggr),%
\label{lam0}%
\end{eqnarray}
\begin{figure}[htp]
\begin{center}
\includegraphics[scale=0.4]{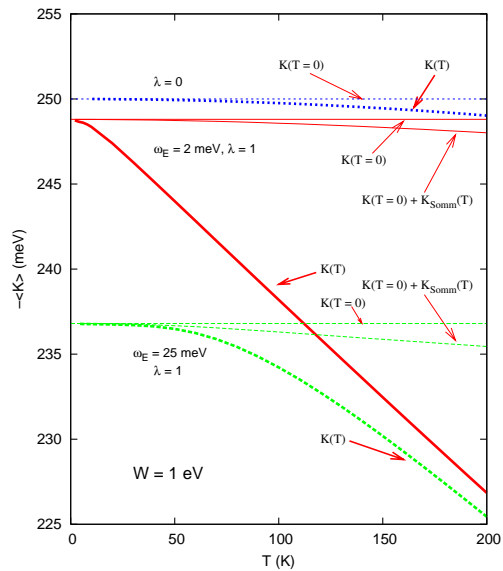}
\caption{ The negative of the kinetic energy vs. temperature showing
the breakdown for the various contributions. The solid (red) curves
are for small boson frequency, $\omega_E = 2$ meV, with $\lambda =
1$. The two thin lines show how the zero temperature value is
modified by the Sommerfeld term by a very small amount; the full
numerical result is given by the thick red curve. A similar
contribution is found for the non-interacting case (actually, there
it is a bit larger) by the dotted (blue) curves. However, here the
relative effect is much larger because only the Sommerfeld term is
responsible for the temperature variation. Finally, for higher boson
frequency, the dashed (green) curves illustrate the amount of
temperature variation due to the Sommerfeld term compared with the
rest.}
\end{center}
\end{figure}
where $f(\omega_E)$ is the Fermi function. This expression is not
very enlightening. However note that the first line is strongly
convergent, so much so that the ${\rm tan}^{-1}$ term is not really
required for most parameter regimes of interest. The Matsubara sum
can then be performed, and one obtains for this first term (without
the $\lambda$ out front), $-\bar{\omega}_E {\rm Im} \psi^\prime({1
\over 2} + i\bar{\omega}_E)$, where $\bar{\omega}_E = \omega_E/(2
\pi T)$ and $\psi(z)$ is the diagamma function; this is the result
one obtains for the infinite bandwidth case. It is positive
definite, starts at unity at zero temperature (so $\lambda_0(T\sim
0) \approx \lambda$) and goes to zero as $(\omega_E/T)^2$ at high
temperature. The other two terms vanish for infinite bandwidth. In
fact the third term is very small for all temperatures with
conventional parameters, while the second term dominates at high
temperature. This is clear since it increases (in magnitude)
linearly with temperature. A plot is shown in Fig. 3. Also shown is
the approximation
\begin{equation}
\lambda^{\rm app}_0(t) \sim -\lambda \bar{\omega}_E {\rm Im}
\psi^\prime({1 \over 2} + i\bar{\omega}_E) + {2 \over
\pi}{\Sigma_2(T) \over W/2},%
\label{lam0app}
\end{equation}%
where we have used Eq. (\ref{sig20}). This approximation is shown by
the symbols and is clearly excellent over the entire temperature
range. The important feature apparent in Fig. 3 is that the mass
enhancement parameter $\lambda_0(T)$ changes sign as the temperature
increases. This is a non-trivial effect; Cappelluti et al.
\cite{cappelluti03} noted a similar phenomenon when impurity
scattering was included (we remind the reader that at high
temperature low frequency phonons behave much like static
impurities). This behaviour is necessary to include to describe the
numerical results accurately, as we shall see below. It is truly
remarkable that at temperatures below $100$ K a modest scattering
($\lambda=1$) off low frequency phonons ($\omega_E=2$ meV) exhibits
a qualitative change (mass enhancement changes sign) when an
electronic cutoff energy (normally taken to be infinite) is
introduced which is 2-3 orders of magnitude larger (1 eV) than the
low energy parameters.

Once we have exactly determined the quasiparticle renormalization
parameter $\lambda_0(T)$ we can estimate the Sommerfeld
contribution. The results are plotted in Fig. 4, where we compare
again the non-interacting case with the $\lambda=1$ cases at two
phonon frequencies. As one can see, in the non-interacting case the
Sommerfeld term, even if small, is the only source of the
temperature variation of the kinetic energy. When interactions are
present instead its contribution cannot account for the temperature
variation of $K(T)$ obtained through the numerical computation of
Eq.\ (\ref{kin1}). We then conclude that to understand the origin of
the strong temperature dependence displayed in Figs.\ 1-2, one must
examine Eq. (\ref{k0}). Note that a different conclusion was instead
drawn in Ref. \onlinecite{toschi05}, where a Hubbard-like
interaction between electrons was considered. Indeed, in this case
the main effect of the interaction is to reduce the effective
quasiparticle bandwidth $1/\lambda_0$, so that as $\lambda_0$
increases the Sommerfeld term continues to provide the largest
contribution to the temperature variation of the kinetic energy.

\section{origin of the linear temperature dependence}

\subsection{Imaginary part of the self energy}

The linear temperature dependence, and, indeed, the strongest
temperature dependence, occurs for low boson frequency, so we focus
on this parameter regime. To put this another way, one should
examine temperatures which are high compared to the boson frequency,
which Fig. 2 illustrates is readily attained in the $100 \sim 200$ K
range when $\omega_E {< \atop \sim} 25$ meV. Then it is clear that
the imaginary part of the self energy at zero frequency increases
linearly with temperature (see Eq. (\ref{sig20})) and one is tempted
to associate the linear temperature dependence observed in $K(T)$
with the linear temperature dependence in $\Sigma_2(T)$. In fact,
Karakozov and Maksimov \cite{karakozov05} noted that the {\em
infinite} frequency limit of the self energy (as defined by the
theory with infinite bandwidth), which we will denote by
$\Sigma_{2\infty}(T)$, also increases linearly with temperature, and
they use this quantity. It is given by
\begin{equation}%
\Sigma_{2\infty}(T) = -\pi {\lambda \omega_E \over 2} {\rm
coth}\bigl({\beta \omega_E \over 2}\bigr).%
\label{s2inf}%
\end{equation}
We should be clear here; the physical self energy always has an
imaginary part which goes to zero beyond the electronic bandwidth
energy scale. However, the standard theory
\cite{allen82,marsiglio03}
eliminates this cutoff, and then Eq. (\ref{s2inf}) provides the
imaginary part of the self energy in the infinite frequency limit.

Either Eq. (\ref{s2inf}) or (\ref{sig20}) provides the correct high
temperature behaviour for the imaginary part of the self energy.
However, as we now illustrate, using this part of the self energy
alone is not sufficient to properly reproduce the numerical result.
To examine what the necessary ingredients are, we proceed in steps,
and illustrate what approximations to the self energy are sufficient
in Eq. (\ref{k0}) to reproduce accurately the numerical result. A
first step (following Ref. \onlinecite{karakozov05}) is to model the
self energy with a constant pure imaginary part, which we take to be
$\Sigma_{2\infty}(T)$. Then the spectral function is given by
\begin{equation}%
A(\epsilon,\omega) = {1 \over \pi}{|\Sigma_{2\infty}(T)| \over
(\epsilon - \omega)^2 + \Sigma_{2\infty}^2(T)}%
\label{speca}%
\end{equation}%
over all frequency, $\omega$. Insertion into Eq. (\ref{k0}) then
allows both integrals to be done analytically. If we first assume
that Eq. (\ref{speca}) holds for all frequencies ($\omega$), then
Eq. (\ref{k0}) results in
\begin{equation}%
K_{0a}(T) = -{W \over 4} \Biggl[ [1 + x^2]{2 \over \pi}{\rm
tan}^{-1} ({1 \over x}) - {2 \over \pi}x \Biggr].%
\label{k0a}%
\end{equation}
where $x \equiv {|\Sigma_{2\infty}(T)| \over W/2}$. Clearly, as
$|\Sigma_{2\infty(T)}|$ decreases compared with $W/2$, $K_0
\rightarrow -W/4$, as for the non-interacting case. This result
(green, dashed curve), along with the numerical result (solid, red
curve), is shown in Fig. 5.
\begin{figure}[tp]
\begin{center}
\includegraphics[scale=0.4]{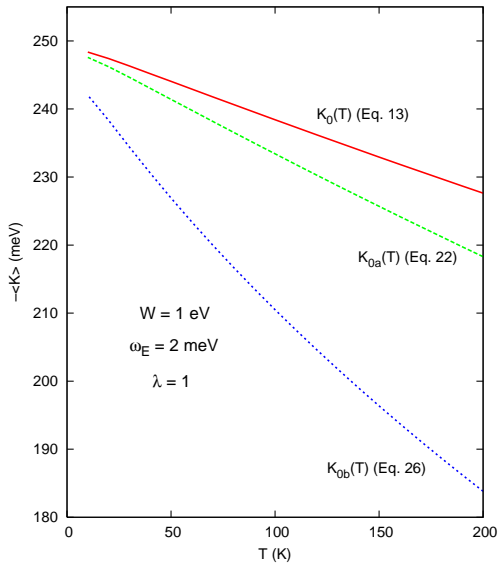}
\caption{ Various approximations of the kinetic energy (without the
Sommerfeld expansion terms) compared with the numerical result
(solid red curve). The green (dashed) curve is given by Eq.
(\ref{k0a}), where only the imaginary part of the self energy is
included. It is not so far off the numerical result. However, the
(relatively small) degree to which a discrepancy exists is
serendipitous, as an obvious improvement (introduction of a cutoff -
see Fig. 6) results in the blue (dotted) curve. This latter
`improved' result is in far worse agreement with the numerics.}
\end{center}
\end{figure}
Note that the numerical result is that for $K_0(T)$; this differs
from the results presented in Figs. 1 - 2 by essentially the
Sommerfeld term, which we leave out of this discussion. In fact we
have checked that a numerical evaluation of $K_0(T)$ based on the
exact self energies inserted into Eq. (\ref{k0}) agrees with the
imaginary axis result with the Sommerfeld term subtracted. The
result in Fig. 5 is somewhat disappointing. Clearly, a linear
dependence on temperature is obtained, and is easily traced to the
linear dependence of $\Sigma_{2\infty}(T) \approx -\pi \lambda T$
for $T>>\omega_E$ (note that $\Sigma_2(T)$ --- see Eq. (\ref{sig20})
--- has the same high temperature behaviour). Indeed, as $x\ll 1$ in
Eq.\ (\ref{k0a}), one finds $K_{0a}(T) \approx
K(0)-(2/\pi)\Sigma_{2\infty}(T)$. However, the slope is evidently
incorrect, and one may wonder how the discrepancy arises.
\begin{figure}[tp]
\begin{center}
\includegraphics[scale=0.4]{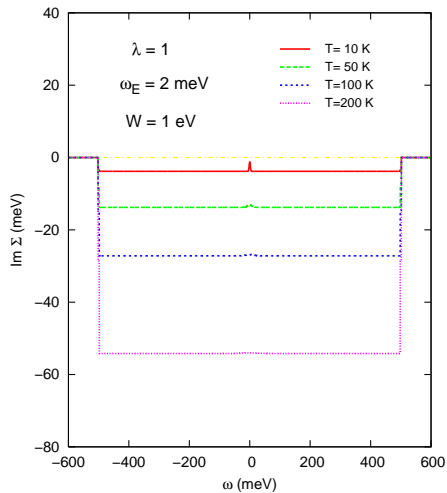}
\caption{ Imaginary part of the self energy for various
temperatures. At high temperatures, the use of a constant to
describe the imaginary part of the self energy works very well;
there is structure at the cutoffs barely visible on the scale shown
here, but they are not very important. Despite this accurate
analytical representation of the imaginary part of the self energy,
this approximation does {\em not} work well for the kinetic energy,
as explained in Fig. 5.}
\end{center}
\end{figure}

A possibility presents itself when one examines more closely the imaginary
part of the self-energy obtained analytically from Eq. (\ref{migdalb}) and
plotted in Fig. 6.  The full expression at positive frequency is:
\begin{eqnarray}
\Sigma_2(z)&=&-\frac{\pi\lambda\omega_E}{2}\left[ \coth(\beta\omega_E/2)
+f(z+\omega_E)-f(z-\omega_E)\right], \nonumber\\
\label{silt}
& &\quad z<W/2-\omega_E, \\
\Sigma_2(z)&=&\frac{\pi\lambda\omega_E}{2}
\left[N(-\omega_E)+f(z-\omega_E)\right], \nonumber\\
\label{siht}
& &\quad W/2-\omega_E<z<W/2+\omega_E, \\
\Sigma_2(z)&=&0, \quad z>W/2+\omega_E.
\end{eqnarray}
As one can see, except for the small energy range below
$z=\omega_E$, $\Sigma_2$ given by Eq.\ (\ref{silt}) can be safely
approximated by the Eq.\ (\ref{s2inf}) above. However, near the band
edge $z=W/2-\omega_E$ the self energy has a first step-like change
to a smaller value, and it then vanishes at $z>W/2+\omega_E$.
Viewed on the scale of Fig. 6, the fine structure present below
$\omega_E$ and near the band edges is perhaps not important, while
accounting for the upper cutoff $z\approx W/2$ clearly should be.
This model can also be solved for analytically. The result is
\begin{eqnarray}
K_{0b}(T) = -{W \over 4} &\Biggl[& [1 + x^2]{2 \over \pi}{\rm
tan}^{-1} ({1 \over x}) - {x \over \pi} {\rm ln}\bigl({4+x^2 \over
x^2}\bigr)\nonumber\\
&-& {x^2 \over \pi} {\rm tan}^{-1} ({2 \over x}) \Biggr],%
\label{specb}%
\end{eqnarray}
where, as before, $x \equiv {|\Sigma_{2\infty}(T)| \over W/2}$. This
result is shown in Fig. 5 as the blue dotted curve. Our attempt at
improvement has resulted in a significantly inferior approximation!

\subsection{Real part of the self energy}

One can attempt to improve approximations to the imaginary part of the self
energy (for example, by using a proper double step at the band edge), but
this does not lead to any improvement in the subsequent $K_0(T)$. Success
is only attained when the real part of the self energy is included. A clear
way to see this is as follows.  One first notes that for an approximation
which invokes a constant (with frequency) for the imaginary part of the
self energy up to the bandwidth (i.e. $\Sigma_2(\omega +i\delta) =
\Sigma_{2\infty}$ for $|\omega|<W/2$, and zero otherwise), then
Kramers-Kronig immediately implies that the real part of the self energy
${\rm Re}\Sigma\equiv \Sigma_1$ is
given by a simple logarithm, i.e.
\begin{equation}%
 \Sigma_1(\omega + i\delta) = {\Sigma_{2\infty}(T) \over
\pi}{\rm ln} \Bigl| {W/2 - \omega \over W/2 + \omega} \Bigr|.%
\label{real_log}
\end{equation}
Then the spectral function can be obtained analytically, and
inserted into Eq. (\ref{k0}), with the spectral function given in general by:
$$
A(\epsilon,\omega) = {1 \over \pi}{|\Sigma_{2}(\omega,T)| \over
(\epsilon - \omega+\Sigma_1(\omega,T))^2 + \Sigma_2^2(\omega,T)}.
$$
The integral over $\epsilon$ can be
performed analytically, but the remaining integral over the
frequency, $\omega$, must be done numerically. When this is done,
the result for the kinetic energy is essentially indistinguishable
from the numerical result at high temperatures.

In light of the `failures' of the approximations in the previous section,
this requires further clarification.  The ($\omega$) frequency integral in
Eq. (\ref{k0}) extends from $-\infty$ to $0$.  In the range $-W/2$ to zero
the self energy has both a real and imaginary part; then the spectral
function is some broad Lorentzian-like function. However, in the range
$-\infty<\omega< -W/2$ the imaginary part of the self energy is zero, so
the spectral function is given by a delta function. The integral over the
variable $\epsilon$ is trivial, and the remaining integral over $\omega$ is
non-zero, {\em because the real part in this frequency range is also
non-zero}. If the real part of the self energy were zero, then this part of
the integral (involving the delta function) is also zero, which is why we
didn't mention it earlier with respect to the approximation $K_{0b}(T)$
(see Eq. (\ref{specb})). Here it must be included, and contributes a
significant part (close to $20\%$ of the total).

This discussion firmly establishes the need to include the real part
of the self energy in an accurate approximation for the kinetic
energy. However, the previous exercise adds little to our fully
numerical results, since one of the integrals must be done
numerically. To see how well a simplified model does, we adopt the
following simpler model. First, we take the imaginary part as
before, cut off at $\pm W/2$. For the real part, if one expands Eq.
(\ref{real_log}) for $|\omega| << W/2$, the result is
\begin{eqnarray}%
\Sigma_1(\omega + i\delta) =& & {2 \over \pi}
|\Sigma_{2\infty}(T)| {\omega \over W/2} \quad {\rm for}
\quad |\omega| < W/2%
\nonumber \\%
= & &{|\Sigma_{2\infty}(T)| \over \pi}{\rm ln} \Bigl| {W/2 + \omega
\over W/2 - \omega} \Bigr| \quad {\rm for}
\quad |\omega| > W/2.\nonumber\\
\label{real_lin}%
\end{eqnarray}
\begin{figure}[tp]
\begin{center}
\includegraphics[scale=0.4]{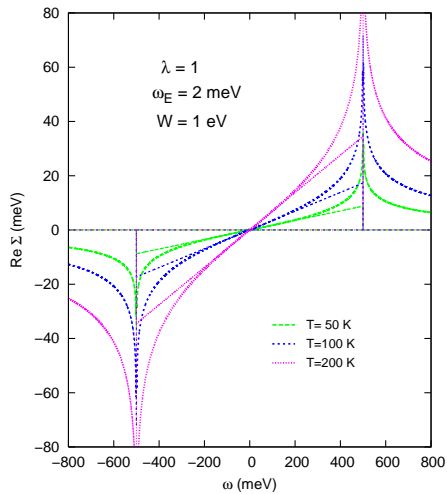}
\caption{ Real part of the self energy for various temperatures.
Note that even at the lowest temperature shown (green, dashed
curve), the overall slope at the origin looks positive on this
scale; in fact there is some structure there, and the slope is in
fact negative, but this plays almost no role in our results. This is
because the most important contributions to the kinetic energy come
from near the bottom of the band. Note that beyond the band edges
the approximation given by Eq. (\ref{real_lin}) is indistinguishable
from the numerical result on this energy scale (a closer look would
reveal some discrepancies --- the numerical result has two
logarithmic divergences at each band edge at low temperature, not
one).}
\end{center}
\end{figure}
In the first part, we have imposed a cutoff at the band edge. Note
that the coefficient of $\omega$ is {\em positive}. This always
occurs at high temperatures, consistent with the negative mass
enhancement displayed in Fig. 4a. In fact, at high temperatures,
$|\Sigma_{2\infty}(T)| \approx \pi \lambda T$, and Eq.
(\ref{real_lin}) reduces to the second term in Eq. (\ref{lam0}),
which dominates at high temperature, as Fig. 4 shows. Notice however
that while approximation (\ref{real_lin}) holds as soon as $T\gtrsim
\omega_E/2$ the change in sign in $\lambda_0(T)$ occurs at higher
temperature. In the second part, we have simply adopted Eq.
(\ref{real_log}). Fig. 7 shows how accurate this approximation is.
In the frequency range $|\omega|<W/2$, the approximation is clearly
crude; beyond this frequency range the approximation given by the
second half of Eq.  (\ref{real_lin}) is indistinguishable from the
numerical result.  (Actually, if we use Eq. (\ref{real_log}), i.e.
the logarithm for the entire frequency range, then this would be
indistinguishable from the numerical result over the {\em entire}
frequency range.  This accounts for our remarks above that insertion
of Eq.  (\ref{real_log}) along with the constant imaginary part
would result in a kinetic energy that is essentially
indistinguishable from the full numerical result.)

Explicitly, Eq. (\ref{k0}) becomes
\begin{eqnarray}
& &K_{0c}(T) = {2 \over W}\int_{-\infty}^{-W/2}d\omega
\int_{-W/2}^{+W/2} d \epsilon \ \epsilon \delta[\omega - \epsilon -
\Sigma_1(\omega+i\delta)]+%
\nonumber \\
& & {2 \over W}\int_{-W/2}^0 d\omega \int_{-W/2}^{+W/2} d
\epsilon \ \epsilon {1 \over \pi} {|\Sigma_{2\infty}(T)| \over
(\epsilon - \omega + \Sigma_1(\omega+i\delta))^2 +
\Sigma_{2\infty}^2(T)}.\nonumber\\
\label{k0c}%
\end{eqnarray}
In the first line, the $\epsilon$ integration is trivial. Care must
be taken concerning the bandedge cutoffs, so that the $\omega$
integration no longer extends to $-\infty$. This lower cutoff,
$\omega_c$, is determined by the condition $\omega_c -
{|\Sigma_{2\infty}(T)| \over \pi}{\rm ln} \Bigl| {W/2 + \omega_c
\over W/2 - \omega_c} \Bigr| = -W/2$, which can easily be solved by
a few iterations. The remaining $\omega$ integral is elementary; we
write it as $K_{0c}^1(T)$:
\begin{equation}%
K_{0c}^1(T) = -{W \over 4} \biggl[ y^2-1 + {2 \over \pi}x \Bigl[
{\rm ln}\bigl({4 \over y^2-1}\bigr)  - y {\rm
ln}\bigl({y + 1 \over y-1}\bigr)\Bigr] \biggr],%
\label{k0c1}%
\end{equation}%
where $x$ is as before, and $y \equiv {|\omega_c| \over W/2}$. The
ratio $y$ is always greater than unity.

\begin{figure}[tp]
\begin{center}
\includegraphics[scale=0.4]{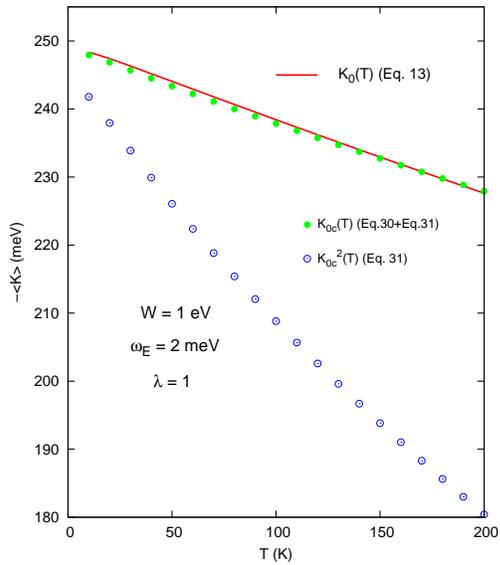}
\caption{ Comparison of $K_{0c}(T)$ approximation (solid green
circles) with numerical results (solid red curve). Agreement is
excellent. Also shown are the results for $K_{0c}^2(T)$ alone, i.e.
without accounting for the delta-function part. Clearly both real
and imaginary parts of the self energy are necessary for accurate
results. }
\end{center}
\end{figure}
In the second line we write $\Sigma_1 = -\tilde{\lambda} \omega$,
where $\tilde{\lambda} = -{2 \over \pi}x$ represents a negative mass
enhancement. Substitution into the integral allows this piece (which
we denote $K_{0c}^2(T)$) to be evaluated analytically as before:
\begin{eqnarray}%
& &K_{0c}^2(T) = -{W \over 4}{1 \over 1 + \tilde{\lambda}} \Biggl[
(1+x^2){2 \over \pi}{\rm tan}^{-1}\bigl({1 \over x}\bigr)\nonumber\\
&-& {x \over
\pi}(1 + \tilde{\lambda}){\rm
ln}\Bigl({(2+\tilde{\lambda})^2 + x^2 \over \tilde{\lambda}^2 + x^2}\Bigr)%
\nonumber \\%
&+& \Bigl(\tilde{\lambda}(\tilde{\lambda} + 2) - x^2\Bigr) {1 \over
\pi} \Bigl({\rm tan}^{-1}\bigl({2 + \tilde{\lambda} \over x}\bigr) -
{\rm tan}^{-1}\bigl({\tilde{\lambda} \over x}\bigr)\Bigr)
\Biggr],\nonumber\\
\label{koc2}%
\end{eqnarray}%
with $x \equiv {|\Sigma_{2\infty}(T)| \over W/2}$. The approximation
$K_{0c}(T)$ consists of the sum of these two pieces, $K_{0c}(T) =
K_{0c}^1(T) + K_{0c}^2(T)$. These are plotted for $\omega_E=2$ meV
and $\lambda = 1$ in Fig. 8 (solid green circles). Agreement with
the numerical result (solid red curve) is now excellent, especially
when one considers the crudeness of the approximation for the real
part of the self energy shown in Fig. 7. We also show the result for
$K_{0c}^2(T)$, i.e. without the correction from the delta-function
piece (which arises only because the real part of the self energy is
non-zero) with the open blue circles. Clearly both real and
imaginary parts of the self energy are required for accurate
results.

\begin{figure}[tp]
\begin{center}
\includegraphics[scale=0.4]{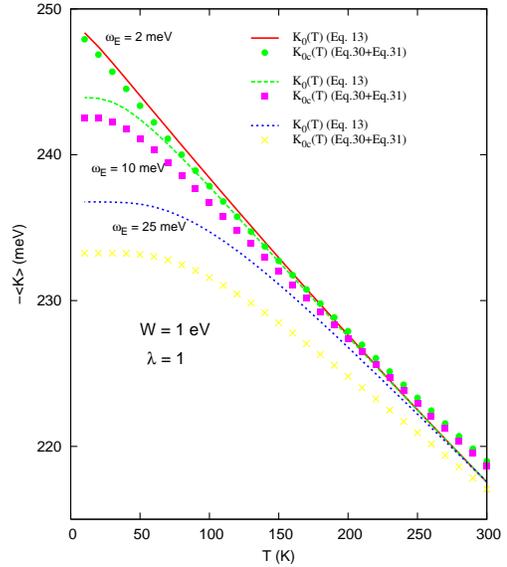}
\caption{ Comparison of analytical model ($K_{0c}(T)$) (symbols)
with numerical results (curves) for several boson frequencies. Not
the expanded vertical scale. Clearly, the model deteriorates as the
boson frequency increases (also note that the result is also no
longer linear). Nonetheless, the discrepancy for $\omega_E = 25$ meV
is still less than $2\%$.}
\end{center}
\end{figure}
Finally, we show results in Fig. 9 for several boson frequencies,
over a somewhat higher temperature range. The approximations used to
obtain $K_{0c}(T)$ clearly deteriorate as the boson frequency
increases with respect to the temperature scale involved. Note in
particular that in the temperature regime where the result is no
longer linear in temperature the qualitative agreement is not so
good (i.e. off by more than a percent). It is also apparent in this
figure that the crude approximation for the real part of the self
energy (first part of Eq. (\ref{real_lin})) results in some minor
discrepancies at higher temperatures. As mentioned before,
replacement of this linear approximation with a single logarithm
(still an approximate result) makes the `analytic' results agree
essentially perfectly with the numerical results.

\section{discussion and summary}

In the previous sections we have described the effects of the
interaction of electrons with a soft collective mode on the
electronic kinetic energy, which is in turn related to the optical
sum rule probed by the experiments. In contrast to the general
expectation for Fermi-liquid systems, here the main contribution to
the temperature variation of the kinetic energy cannot be attributed
to the Sommerfeld term, i.e. to the thermal smearing of the Fermi
occupation number. Instead, the relevant effects come from the
temperature dependence set by the interaction itself, via the
electronic self energy. While discussing the sum rule behavior the
presence of a finite bandwidth $W$ is a crucial ingredient, since
the temperature variations are expected to vary as a power of $T/W$
or $\Sigma_2/W$. The latter case is relevant here, with the kinetic
energy, as given by Eq.s\ (\ref{k0c1})-(\ref{koc2}), scaling as
$\Sigma_2/W$. Moreover, since $\Sigma_2(T)=-\pi\lambda T$ at high
temperatures, we also found a linear temperature depletion of the
optical sum. However, in deriving this result we had to properly
take into account also the structure of the real part of the self
energy, whose behavior at low frequency is quite strongly affected
by the presence of a finite-bandwidth cut-off, in contrast to the
general expectation coming from the analysis of the
infinite-bandwidth approximation.\cite{allen82,marsiglio03}

Finally, we would like to present a simple argument which can help the
reader to understand the difference between the effect of thermal smearing
of the Fermi function with respect to the effect of the interaction
considered here. Going back to the expression (\ref{kinetic_energy}) of
the kinetic energy, we can understand the Sommerfeld term as due to the
fact that by varying the temperature one has a variation of the occupation
factor $\delta n(\xi)$ of order $\xi/T$ over a range $\sim T$ around the
Fermi level (here $\xi=\epsilon-\mu$) and $0$ otherwise so that:
\begin{equation}
\delta K\sim \frac{1}{W}\int_{-T}^{T} d\xi \, (\xi+\mu) \,
\delta n(\xi)\sim \frac{T^2}{W}=W\left(\frac{T}{W}\right)^2.
\end{equation}
As a consequence, the $T^2$ power is due to the fact that the
integration of energy, which gives $\epsilon^2$, on a range $T/W$,
produces a power of $(T/W)^2$ in the kinetic energy. The question
then arises: do interactions similarly simply change the occupation
number in a region $|\Sigma_2|/W$ around the Fermi level, which
would then lead to a $(|\Sigma_2|/W)^2$ dependence of the kinetic
energy for the interacting case? The answer we have found is
clearly, no. Indeed, when the occupation number coming from Eq.\
(\ref{momentum_distribution}) is considered, using the same
approximations (\ref{s2inf}) and (\ref{real_lin}) considered above
for the self-energy we can estimate $n_0(\epsilon)$ (i.e. the
contribution coming from the $T=0$ limit of the Fermi function, in
analogy with the definition (\ref{k0}) of $K_0(T)$) as:
\begin{equation}
n_0(\epsilon)=\frac{1}{\pi(1+\tilde\lambda)}
\left[\tan^{-1}\frac{W(1+\tilde\lambda)/2+
\epsilon}{|\Sigma_{2\infty}|}-\tan^{-1} \frac{\epsilon}{|\Sigma_{2\infty}|}
\right],
\end{equation}
As a consequence, one can roughly say that $n_0(\epsilon)\sim 1-\alpha
{|\Sigma_{2\infty}|}/W$, with some coefficient $\alpha$ of order 1, in most
of the energy range between $-W/2$ and $0$, and not only in a small region
around the Fermi level. In other words, the interaction affects the
occupation number even far from the Fermi level, so that the energy
integration in Eq.\ (\ref{kinetic_energy}) is now scaling as:
\begin{equation}
\delta K\sim \frac{1}{W}\int_{-W/2}^{\mu} d\epsilon \, \epsilon \,
\alpha\frac{|\Sigma_{2\infty}|}{W/2}\sim
\alpha W\left(\frac{|\Sigma_{2\infty}|}{W}\right).
\label{naive}
\end{equation}
Thus, the kinetic energy is directly proportional to the temperature
variation of the imaginary part of the self energy, even though the
exact coefficient $\alpha$ should be determined taking care also of
contributions from the real part of the self-energy. Even though the
previous relation is not valid for stronger interactions (as in
Hubbard-like models \cite{toschi05}), one can reasonably expect that
by modeling the inelastic scattering with a more general phonon-like
spectrum, the sum rule behavior will roughly follow the modification
introduced in the electronic self-energy in the way indicated by
Eq.\ (\ref{naive}). This possibility leads to an understanding of
the experimental observation of a non-quadratic temperature
dependence of the sum rule in some cuprate superconductors, where
also the high-energy scattering rate shows a linear temperature
scaling \cite{hwang,bontemps06}.

\begin{acknowledgments}
We wish to thank E. Cappelluti, C. Castellani, A. Toschi, D. van der
Marel, and A. Millis for helpful discussions. In addition the
hospitality of the Department of Condensed Matter Physics at the
University of Geneva is greatly appreciated. This work was supported
in part by the Natural Sciences and Engineering Research Council of
Canada (NSERC), by ICORE (Alberta), by the Canadian Institute for
Advanced Research (CIAR), and by the University of Geneva.
\end{acknowledgments}

\bibliographystyle{prb}

\begin{thebibliography}{1}

\bibitem{basov99}
D.N. Basov, S.I. Woods, A.S. Katz, E.J. Singley, R.C. Dynes, M. Xu,
D.G. Hinks, C.C. Homes, and M. Strongin, Science {\bf 283}, 49
(1999); A.S. Katz, S.I. Woods, E.J. Singley, T. W. Li, M. Xu, D.G.
Hinks, R.C. Dynes and D.N. Basov, Phys. Rev. B{\bf 61}, 5930 (2000).

\bibitem{molegraaf02}  H.J.A.~Molegraaf, C.~Presura, D.~van~der~Marel,
P.H.~Kes, and M.~Li, Science {\bf 295}, 2239 (2002).

\bibitem{santander-syro03}
A.F.~Santander-Syro, R.P.S.M.~Lobo, N.~Bontemps, Z.~Konstantinovic,
Z.Z.~Li, H.~Raffy, Europhys. Lett. {\bf 62}, 568 (2003);
A.F.~Santander-Syro, R.P.S.M.~Lobo, N.~Bontemps, W.~Lopera,
D.~Girata, Z.~Konstantinovic, Z.Z.~Li, H.~Raffy, Phys.~Rev.~B {\bf
70}, 134504 (2004). G.~Deutscher, A.F.~Santander-Syro,  N.~Bontemps,
Phys. Rev. B {\bf 72}, 092504 (2005).


\bibitem{homes} C.C.~Homes,
S.V.~Dordevic, D.A.~Bonn, R.~Liang, W.N.~Hardy, Phys.~Rev.~B {\bf
69}, 024514 (2004).

\bibitem{boris} A.V.~Boris, N.N.~Kovaleva, O.V.~Dolgov,
  T.~Holden, C.T.~Lin, B.~Keimer, C.~Bernhard, Science {\bf 304}, 708
  (2004).

\bibitem{ortolani} M.~Ortolani, P.~Calvani, S.~Lupi, \prl {\bf 94},
  067002 (2005).

\bibitem{hwang} J.~Hwang, J.~Yang, T.~Timusk, S.~G.~Sharapov,
  J.~P.~Carbotte,  D.~A.~Bonn, Ruixing Liang, and W.~N.~Hardy, \prb {\bf 73},
 014508 (2006).

\bibitem{bontemps06} N. Bontemps, R.P.S.M. Lobo, A.F.
Santander-Syro, and A. Zimmers, cond-mat/0603024.

\bibitem{hirsch92} J.E. Hirsch, Physica C 199, 305 (1992); Physica C 201, 347
(1992).

\bibitem{hirsch00} J.E. Hirsch and F. Marsiglio, Physica C {\bf 331}, 150
(2000); J.E. Hirsch and F. Marsiglio, Phys. Rev. B62, 15131 (2000).

\bibitem{norman02}
M.R. Norman and C. P\'epin, Phys. Rev. B {\bf 66}, 100506 (2002).

\bibitem{karakozov02} A.E. Karakozov, E.G. Maksimov, and O.V. Dolgov,
Solid State Commun. {\bf 124}, 119 (2002).


\bibitem{schachinger05} E. Schachinger and J. P. Carbotte, Phys. Rev. B
{\bf 72}, 014535 (2005).


\bibitem{marsiglio06} F. Marsiglio, Phys. Rev. B {\bf 73}, 064507 (2006).


\bibitem{knigavko04} A. Knigavko, J. P. Carbotte, and F.
Marsiglio, Phys. Rev. B {\bf 70}, 224501 (2004); Europhysics Letts.
{\bf 71}, 776 (2005).

\bibitem{benfatto05}
L.~Benfatto, S.~Sharapov, and H.~Beck, Eur. Phys. J. B {\bf 39}, 469
(2004); L. Benfatto, S. G. Sharapov, N. Andrenacci, and H. Beck, Phys. Rev.
B {\bf 71}, 104511 (2005).


\bibitem{knigavko05} A. Knigavko and J.P. Carbotte, Phys. Rev. B {\bf 72},
  035125 (2005); cond-mat/0602681, to be published in Physical Review B.

\bibitem{maier04}
Th. A. Maier, M. Jarrell, A. Macridin, and C. Slezak, Phys. Rev.
Lett. {\bf 92}, 027005 (2004).

\bibitem{toschi05}
A. Toschi, M. Capone, M. Ortolani, P. Calvani, S. Lupi, and C. Castellani
Phys. Rev. Lett. {\bf 95}, 097002 (2005).

\bibitem{haule06}
K. Haule and G. Kotliar, cond-mat/0601478.

\bibitem{benfatto06}
L.~Benfatto and S.~Sharapov, cond-mat/0508695, to appear in J. Low
Temp. Phys. (2006).

\bibitem{vandermarel03} D. van der Marel,  H.J.A. Molegraaf, C. Presura,
and I. Santoso, in {\it Concepts in Electron Correlations}, edited
by A. Hewson and V. Zlatic (Kluwer, 2003). See also
cond-mat/0302169.


\bibitem{lanzara01} A. Lanzara, P.V. Bogdanov, X.J. Zhou, S.A. Kellar,
D.L. Feng, E.D. Lu, T. Yoshida, H. Eisaki, A. Fujimori, K. Kishio,
J.-I. Shimoyama, T. Noda, S. Uchida, Z. Hussain, and Z.-X. Shen,
Nature (London) {\bf 412}, 510 (2001).

\bibitem{chubukov05} See, for example,
A. V. Chubukov and J. Schmalian, Phys. Rev. B {\bf 72}, 174520
(2005).

\bibitem{karakozov05} A.E. Karakozov and E.G. Maksimov, cond-mat/0511185.

\bibitem{maldague77} P.F. Maldague, Phys. Rev. B{\bf 16}, 2437
(1977).

\bibitem{remark1} We have previously discussed possible temperature dependence
arising from band edge effects \cite{benfatto05,marsiglio06}. In
what follows here we will assume particle-hole symmetry, so that
there is no "extra" temperature dependence arising from these
effects.

\bibitem{ashcroft76} N.W. Ashcroft and N.D. Mermin, In: {\it Solid State
Physics} (Saunders College Publishing, New York 1976)p. 518.


\bibitem{dogan03} F. Dogan and F. Marsiglio, Phys. Rev. B {\bf 68}, 165102
(2003).

\bibitem{cappelluti03} E. Cappelluti and L. Pietronero, Phys.
Rev. B {\bf 68}, 224511 (2003).


\bibitem{allen82} P.B. Allen and B. Mitrovi\'{c},
In: {\it Solid State Physics}, edited by H. Ehrenreich, F.~Seitz,
and D. Turnbull (Academic, New York, 1982) Vol. 37, p.1.

\bibitem{marsiglio03} F. Marsiglio and J.P. Carbotte, Review Chapter in
`The Physics of Conventional and Unconventional Superconductors'
edited by K.H. Bennemann and J.B. Ketterson (Springer-Verlag), pp.
233-345 (2003). See also cond-mat/0106143.

\end{thebibliography}

\end{document}